\documentclass[onecolumn,showpacs,preprintnumbers,amsmath,amssymb]{revtex4}
\usepackage{graphicx}
\usepackage{dcolumn}
\usepackage{here}
\usepackage{bm}

\begin{document}

\title{Quantum mechanism of Biological Search}

\author{Younghun Kwon}
\email{yyhkwon@hanyang.ac.kr}
\affiliation{Department of Physics, Hanyang University, \\Ansan, Kyunggi-Do, 425-791, South Korea}

\date{\today}

\begin{abstract}
We wish to suggest an algorithm for biological search including DNA search. 
Our argument supposes that biological search be performed by quantum search.If we assume this, 
we can naturally answer the following long lasting puzzles such that "Why does DNA use the helix structure?" and "How can the evolution in biological system occur?".

\end{abstract}

\maketitle

We wish to suggest an algorithm for biological search including DNA search. This algorithm has novel features which are considerable biological interest.
 Algorithms for biological search have already been proposed by several groups. One of them, based on quantum search \cite{1}, is specially noteworthy because it could explain  
the magic numbers such as 4(number of codon), 10(number of tRNA), 20(number of amino acid) etc, which appear in biological systems\cite{2}. Even though Patel could explain the magic number by quantum search argument, he failed to give the detailed mechanism for biological search by quantum search.\\   
 We wish to put forward a radically different mechanism for biological search. Our argument supposes that biological search be performed by quantum search.If we assume this, we can naturally answer the following two long lasting puzzles such that "Why does DNA use the helix structure" and "How can the evolution in biological system occur?". Our argument starts with quantum search hamiltonian given by\cite{3}

\begin{eqnarray}
H=E(|w\rangle \langle w|+|\psi\rangle \langle \psi|)+\epsilon(e^{i\phi}|w\rangle \langle \psi|+e^{-i\phi}|\psi \rangle \langle w|) \nonumber \\
\end{eqnarray}

Here $\epsilon$($\leq E$) is a constant in unit of energy and $\phi$ is a phase.We use $|w\rangle$ and $|\psi\rangle$ to denote the target and the initial states, respectively.The merit of above Hamiltonian is that when the unitary operation $e^{-iHt}$(we put $\hbar=1$ throughout) applies to the states, the initial state moves to the target one.

If we fix the phase $\phi=0$, then the hamiltonian becomes

\begin{eqnarray}
H=E(|w\rangle \langle w|+|\psi\rangle \langle \psi|)+\epsilon(|w\rangle \langle \psi|+|\psi \rangle \langle w|) \nonumber \\
\end{eqnarray}

Suppose that the target state $|w\rangle$ might get the phase $e^{i \eta} |w\rangle$($|w\rangle \rightarrow e^{i \eta} |w\rangle$). Then the Hamiltonian becomes

\begin{eqnarray}
H=E(|w\rangle \langle w|+|\psi\rangle \langle \psi|)+\epsilon(e^{i\eta}|w\rangle \langle \psi|+e^{-i\eta}|\psi \rangle \langle w|) \nonumber \\
\end{eqnarray}

  When the phase $\eta$ is not $n \pi$, by the analysis of the above hamiltonian (3),
the probability to find target by unitary evolution will not be 1\cite{4}. This implies that the phase $\eta$ is  important role in biological search. Here we can ask how the biological system can get the phase $\eta$. In DNA, the natural way to get the phase is to take the helix structure. When DNA moves through the helix structure, the target state naturally gets the phase. If it is true, we can check this argument by an experiment that if the phase of DNA can be changed, the DNA will not couple with A and T or C and G with a small possibility. And it might be one of the reason of errors in DNA reproduction.\\
Another suprising feaure is that when the phase $\eta$ is not $n \pi$, there might be a speedup in searching the target\cite{4}. That is, when the phase $\eta$ is not $n \pi$ and satisfies some condition, even though the biological system cannot find the target with perferction, the searching ability becomes stronger than when the phase $\eta$ is  $n \pi$. The fact that the searching ability improves means that the biological system can fit more suitably the changing environment. This ability can explain naturally how the biological system could evolve through the years. In other words, we might say that the small errors in quantum system can cause biological evolution. 

\section*{Acknowledgement}
Y. Kwon thanks the theory group of University of Rochester for their hospitality. 
Y. Kwon is supported by Hanyang University.


\begin{thebibliography}{99}


\bibitem{1} L. Grover, Phys. Rev. Lett. $\mathbf{79}$, 325(1997)

\bibitem{2} A. Patel, Pramana $\mathbf{56}$,365(2001) 

\bibitem{3} J. Bae and Y. Kwon, Phys. Rev. A $\mathbf{66}$, 012314(2002)

\bibitem{4} J. Bae, Y. Kwon, I.Bak and D. Yoon, preprint to appear

\end{thebibliography}
\end{document}